\documentstyle[onecolumn,epsfig]{mn}

\begin{document}

\title{Effects of quintessence on observations of Type Ia SuperNovae in the
clumpy Universe}

\author[M. Sereno, E. Piedipalumbo, M.V. Sazhin]
{M. Sereno,$^{1,2}$\thanks{E-mail: sereno@na.infn.it.} E.
Piedipalumbo,$^{1,2}$ M.V. Sazhin$^{3,4}$
\\$^1$ Dipartimento di Scienze Fisiche, Universit\`{a} degli Studi di Napoli
``Federico II", \\ Comp. Univ. Monte S. Angelo, Ed. G, Via Cinthia,
80126, Napoli, Italia.
\\$^2$ Istituto Nazionale di Fisica Nucleare,
sez. Napoli,\\ Comp. Univ. Monte S. Angelo, Ed. G, Via Cinthia, 80126,
Napoli, Italia.
\\$^3$ Sternberg Astronomical Institute, Moscow, Russia.
\\$^4$ Osservatorio Astronomico di Capodimonte, Via Moiariello, 16, 80131
Napoli, Italia.}
\maketitle

\begin{abstract}
We discuss the amplification dispersion in the observed luminosity of
standard candles, like supernovae (SNe) of type Ia, induced by
gravitational lensing in a Universe with dark energy (quintessence).
We derive the main features of the magnification probability
distribution function (pdf) of SNe in the framework of on average
Friedmann-Lema\^{\i}tre-Robertson-Walker (FLRW) models for both lensing by
large-scale structures and compact objects. Analytic expressions, in
terms of hypergeometric functions, for luminosity distance--redshift
relations in a flat Universe with homogeneous dark energy have been
corrected for the effects of inhomogeneities in the pressureless dark
matter (DM). The magnification pdf is strongly dependent on the
equation of state, $w_Q$, of the quintessence. With no regard to the
nature of DM (microscopic or macroscopic), the dispersion increases
with the redshift of the source and is maximum for dark energy with
very large negative pressure; the effects of gravitational lensing on
the magnification pdf, i.e. the mode biased towards de-amplified
values and the long tail towards large magnifications, are reduced for
both microscopic DM and quintessence with an intermediate $w_Q$.
Different equations of state of the dark energy can deeply change the
dispersion in amplification for the projected observed samples of SNe
Ia by future space-born missions. The ``noise" in the Hubble diagram
due to gravitational lensing strongly affects the determination of the
cosmological parameters from SNe data. The errors on the pressureless
matter density parameter, $\Omega_M$, and on $w_Q$ are maximum for
quintessence with not very negative pressure. The effect of the
gravitational lensing is of the same order of the other systematics
affecting observations of SNe Ia. Due to the lensing by large-scale
structures, in a flat Universe with $\Omega_M =0.4$, at $z=1$ a
cosmological constant ($w_Q=-1$) can be interpreted as dark energy
with $w_Q <-0.84$ (at $2$-$\sigma$ confidence limit).
\end{abstract}

\section{Introduction}
During the last years, two independent groups, the High-$z$ SuperNova
Search Team \cite{sm&al98} and the Supernova Cosmology Project
\cite{pe&al99} have given strong evidences of the acceleration of the
Universe's expansion \cite{ri&al98,pe&al99}. Several other
observational and theoretical evidences, like  measurements of the
anisotropy of the Cosmic Microwave Background Radiation
\cite{deb&al00} and inflationary theories, strongly support a flat or
nearly flat Universe. On the other hand, direct measurements of
$\Omega_M$ from dynamical estimates or X-ray and lensing observations
of clusters of galaxies indicates that $\Omega_M$ is significantly
less than unity, $\Omega_M \simeq 0.3$ \cite{tur00}. To solve this
puzzle, a new type of energy component in the Universe, now called
dark energy or quintessence, was proposed; dark energy with strongly
negative pressure is required to explain acceleration ($w_Q\equiv p_Q
/\rho_Q <-1/3$, where $p_Q$ and $\rho_Q$ are, respectively, the
pressure and energy density of the dark energy).

Observations of SNe Ia, which at low redshifts are sensitive to the
deceleration parameter $q_0=(\Omega_M + (1+3w_Q)\Omega_Q)/2$ where
$\Omega_Q$ is the density parameter of the dark energy, rely on
several properties of these sources. SNe Ia are very luminous and have
a small intrinsic dispersion in their peak absolute magnitude, $\delta
M
\stackrel{<}{\sim} 0.3$ \cite{fi&ri99}. These features make them the long expected
standard candles for cosmology.

A standard candle is a source with known intrinsic luminosity ($L$)
(or absolute magnitude). Measurements of its apparent flux
(${\cal{F}}$) allow us to determine the photometric distance $D_L$ to
the source via equation
\begin{equation}
\label{eq:1}
D_L \equiv \sqrt{\frac{L}{4\pi {\cal{F}}}}.
\end{equation}
Using standard candles, it is possible to plot the Hubble diagram
(that is, the redshift of an object versus cosmological distance to it
or vice versa) with very high precision and estimate the global
cosmological parameters.

There are several candidates for the dark energy. The oldest one,
initially introduced by Albert Einstein as a new fundamental constant
of nature, is the cosmological constant ($w_Q=-1$). After the
formulation of inflationary theory, cosmologists found that a
$\Lambda$ term can be introduced dynamically
\cite{dol90,zel92,sa&st00}; a dynamical $\Lambda$ term by a scalar
field slowly rolling down its potential ($w_Q \geq
-1$) \cite{pe+ra88,wett88,os&al95,ca&al98,zl&al99,rit&al00,rub00,ru&sc01}
can support a static energy component with positive energy density but
negative pressure. Other possibilities for the quintessence are
represented either by networks of light, non-intercommuting
topological defects \cite{vi84,sp+pe97} ($w_Q
=-m/3$, where $m$ is the dimension of the defect: for a string  $w_Q=-1/3$;
for a domain wall $w_Q=-2/3$) or by the so called $X$-matter
\cite{ch&al97,tu&wh97}. Alternatively to quintessence, a Universe
filled with Chaplygin gas \cite{ka&al01} is an additional alternative
to obtain a negative pressure. Generally, the equation of state $w_Q$
evolves with the redshift, and the feasibility of reconstructing its
time evolution has been investigated
\cite{co&hu99,ch&na00,ma&al00,sa&al00,gol+al01,hu&tu01,na&ch01,pa+al01,wa&ga01,wa+lo01,we+al01a,we+al01b,ya&fu01,ge+ef02}.
Since in flat FLRW models the distance depends on $w_Q$ only through a
triple integral on the redshift \cite{ma&al00}, $w_Q (z)$ can be
determined only provided a prior knowledge of the matter density of
the Universe \cite{gol+al01}. In what follows, we will consider only
the case of a constant equation of state.

Astrophysical sources other than SNe have been long investigated to
build the Hubble diagram. Two independent luminosity estimators, the
first one based on the variability of Gamma-Ray Bursts (GRBs)
\cite{rei01a,rei01b} and the second one derived from the time lag
between peaks in hard and soft energies \cite{nor+al00}, have been
recently proposed to infer the luminosity distance to these sources.
On the other hand, standard rods, as compact radio sources
\cite{gur+al99} or double radio galaxies \cite{gue+al00}, have been
long studied to evaluate the angular diameter distance to cosmological
sources. With no regard to their different physical origins, all these
observations are affected by gravitational lensing of the sources. In
this paper, we want to quantify the effect of inhomogeneities in the
pressureless matter on the determination of the distance. In
particular, we will study SNe Ia, whose importance in the
determination of the cosmological parameters makes necessary a
complete study of all systematics. For light beams propagating in the
inhomogeneous Universe, the expression of the luminosity distance in
terms of the cosmological parameters, as obtained from
Eq.~(\ref{eq:1}), changes with respect to the corresponding FLRW
model. Mathematical considerations for non-flat models of Universe are
done in Sereno et al.
\shortcite{ser+al01}. In this paper, we will discuss the simple case of the flat
Universe and the influence of clumpiness on the Hubble diagram. We
consider inhomogeneous pressureless matter and smooth dark energy. For
the more general case of inhomogeneous quintessence see Linder
\shortcite{li88} and Sereno et al.
\shortcite{ser+al01}.

The paper is as follows. In Sect. 2, we introduce the on average FLRW
models; we discuss the Dyer-Roeder (DR) equation and its analytical
solution in terms of hypergeometric functions. In Sect. 3, we consider
the case of the homogeneous Universe. Section 4 contains the
discussion on the statistical nature of the lensing dispersion induced
either by large-scale structures or compact objects; we study the
magnification pdf induced by gravitational lensing and the connected
systematic errors in the estimate of $\Omega_M$ and $w_Q$. Section 5
is devoted to some final considerations.

\section{The luminosity distance--redshift relation}

The standard Hubble diagram is computed with relations that hold in
FLRW models, that is assuming all gravitating pressureless matter
homogeneously distributed. Instead, observational data are taken in
the inhomogeneous Universe, where sources most likely appears to be
de-magnified relative to the standard Hubble diagram. In fact, light
bundles, propagating far from clumps along the line of sight from the
source to the observer, experience a matter density less than the
average matter density of the Universe. Several attempts have
addressed the problem of the redshift dependence of the distance in a
clumpy Universe: by relaxing the hypothesis of homogeneity and using
the Tolman-Bondi metric instead of the FLRW one \cite{cele00}; by
quantifying the small deviations from the isotropy and homogeneity of
the Ricci scalar \cite{tren01}; by considering a local void
\cite{tomi01}. One of the historically more important and widely used
framework is the on-average FLRW Universe \cite{sef,se+al94}, where it
is assumed that the relations on a large scale are the same of the
corresponding FLRW Universe, while inhomogeneities only affect local
phenomena like the propagation of light. A simple way to account for
this scenario is to introduce the smoothness parameter $\alpha_M$,
that, in a phenomenological way, represents the magnification effect
experienced by the light beam \cite{tomi98,wan99}. In general, the
smoothness parameter is redshift dependent since the degree of
smoothness in pressureless matter evolves with time. When
$\alpha_M=1$, we reduce to the FLRW case; when $\alpha_M<1$, there is
a defocusing effect; $\alpha_M=0$ represents a totally clumped
Universe. This limiting case, sometimes known as ``empty-beam
approximation", corresponds to the maximum distance to a source for
light bundles which have not passed through a caustic \cite{sef}.
Historically, $\alpha_M$ is defined as the fraction of pressureless
matter smoothly distributed. The distance recovered in on average FLRW
models, sometimes known as DR distance, has been long studied
\cite{ze64,ka69,dy&ro72,dy&ro73,kor86,li88,se+al94} and now is
becoming established as a very useful tool for the interpretation of
experimental data \cite{kan98,ka&th00,pe&al99,gi&al00}. The DR
equation for the angular diameter distance, $D_A$, in the case of a
Universe with inhomogeneous pressureless matter and dark energy with
constant equation of state, has already been obtained starting from
the optical scalar equations \cite{li88,gi&al00} or using the multiple
lens-plane theory \cite{ser+al01}. For flat Universes, $\Omega_M
+\Omega_Q
=1$, it is
\begin{equation}
\label{dr1}
(1+z)^2
\left(1+\mu(1+z)^{3w_Q}\right)\frac{d^2D_A}{dz^2}+\frac{(1+z)}{2}\left(
7+(3w_Q +7)\mu(1+z)^{3w_Q} \right)\frac{d D_A}{dz}
\end{equation}
\[
 +\frac{3}{2}
\left( \alpha_M + (w_Q +1)\mu (1+z)^{3w_Q}
\right)D_A=0,
\]
where $\mu \equiv \frac{\Omega_Q}{\Omega_M}$ and $\alpha_M$ is the
smoothness parameter. The boundary conditions on Eq.~(\ref{dr1}) are
\begin{eqnarray}
\label{dr2}
D_A(0)& = & 0, \\
\left. \frac{d}{dz}D_A \right|_{z=0} & = & \displaystyle{\frac{c}{H_0}}.\nonumber
\end{eqnarray}
It is straightforward to obtain the corresponding equation for the
luminosity distance. Using the Etherington principle \cite{et33},
\begin{equation}
\label{dr3}
D_L=(1+z)^2D_A,
\end{equation}
we can substitute in Eq.~(\ref{dr1}),
\begin{equation}
\label{dr4}
(1+z)^2(1+\mu(1+z)^{3w_Q})\frac{d^2 D_L}{dz^2}-\frac{(1+z)}{2}\left( 1
+(1
-3w_Q)\mu(1+z)^{3w_Q} \right)\frac{d D_L}{dz}
\end{equation}
\[
+ \left(
\frac{3\alpha_M -2}{2}+ \frac{1-3w_Q}{2}\mu (1+z)^{3w_Q}
\right)D_L=0;
\]
the boundary conditions are, again,
\begin{eqnarray}
\label{dr5}
D_L(0)& = & 0, \\
\left. \frac{d}{dz}D_L \right|_{z=0} & = & \displaystyle{\frac{c}{H_0}}.\nonumber
\end{eqnarray}
The solution of Eq.~(\ref{dr4}), satisfying the boundary conditions in
Eq.~(\ref{dr5}), takes the form
\begin{equation}
\label{dr6}
D_L(z)=\frac{c}{H_0}\frac{D_1(0)D_2(z)-D_1(z)D_2(0)}{W(0)},
\end{equation}
where $D_1(z)$ and $D_2(z)$ are two linearly independent solutions of
Eq.~(\ref{dr4}) and $W(z)\equiv D_1(z)\frac{D_2(z)}{dz}-\frac{d
D_1(z)}{dz}D_2(z)$ is the Wronskian of the solutions system. In what
follows, we will consider the DR equation with a constant smoothness
parameter. We introduce the parameter
\begin{equation}
\beta \equiv \frac{\sqrt{25-24\alpha_M}}{4}.
\end{equation}
To solve Eq.~(\ref{dr4}), we perform the transformation of both the
independent and dependent variables,
\begin{equation}
\label{dr7}
u \equiv -\mu (1+z)^{3w_Q},\ \ D_L(z)\equiv
u^{\frac{3+4\beta}{12w_Q}}R_L(z).
\end{equation}
With such a transformation, Eq.~(\ref{dr4}) reduces to the
hypergeometric equation for $R_L$,
\begin{equation}
\label{dr8}
\frac{d^2R_L}{du^2}+\left[\left( 1+ \frac{2\beta}{3w_Q}\right)\frac{1}{u}-\frac{1}{2(1-u)} \right]
\frac{d R_L}{du}-\left( \frac{4\beta-1}{12w_Q}\right)
\left( \frac{4\beta +1}{12w_Q}+\frac{1}{2} \right) \frac{1}{u(1-u)}R_L=0.
\end{equation}
A pair of independent solutions of Eq.~(\ref{dr8}) is
\begin{eqnarray}
\label{dr9}
& & R_1(u)=\ _2F_1\left[
\frac{4\beta-1}{12w_Q},\frac{4\beta +1}{12w_Q}+\frac{1}{2},\frac{2\beta}{3w_Q}+1,u\right],
\\
& & R_2(u)=u^{-\frac{2\beta}{3w_Q}} \ _2F_1\left[
-\frac{4\beta+1}{12w_Q},\frac{-4\beta+1}{12w_Q}+\frac{1}{2},-\frac{2\beta}{3w_Q}+1,u\right],
\nonumber
\end{eqnarray}
where $_2F_1$ is the hypergeometric function of the second type.
Inserting the expressions for $R_1$ and $R_2$ in Eq.~(\ref{dr7}) and
substituting in Eq.~(\ref{dr6}), we have the final expression for the
luminosity distance,
\begin{eqnarray}
\label{dr10}
\lefteqn{D_L(z)=\frac{c}{H_0}\frac{1}{2\beta \sqrt{\Omega_M}}} \\
& &
{\times} \left\{ (1+z)^{\frac{3}{4}+\beta}\
_2F_1\left[-\frac{4\beta +1}{12w_Q}, \frac{1}{2}+\frac{1-4\beta}{12w_Q},
1-\frac{2\beta}{3w_Q},\frac{\Omega_M-1}{\Omega_M} \right] \right.
\nonumber
\\ & &
\
{\times} _2F_1\left[\frac{4\beta -1}{12w_Q}, \frac{1}{2}+\frac{4\beta
+1}{12w_Q},1+\frac{2\beta}{3w_Q},
\frac{\Omega_M-1}{\Omega_M}(1+z)^{3w_Q} \right]  \nonumber \\
& &
- (1+z)^{\frac{3}{4}-\beta}\
_2F_1\left[-\frac{4\beta +1}{12w_Q}, \frac{1}{2}+\frac{1-4\beta}{12w_Q},
1-\frac{2\beta}{3w_Q},\frac{\Omega_M-1}{\Omega_M}(1+z)^{3w_Q}
\right] \nonumber  \\
& & \left. {\times} \
_2F_1\left[\frac{4\beta -1}{12w_Q}, \frac{1}{2}+\frac{4\beta +1}{12w_Q},1+\frac{2\beta}{3w_Q},
\frac{\Omega_M-1}{\Omega_M} \right]\right\}. \nonumber
\end{eqnarray}

For the case of a cosmological constant, $w_Q=-1$, Eq.~(\ref{dr10})
reduces to equation~(16) in Kantowski \& Thomas \shortcite{ka&th00},
as we can see by using the property of the hypergeometric functions
\begin{equation}
_2F_1\left[a,b,c,x\right]=\frac{1}{(1-x)^a}\ _2F_1\left[ a,c-b,c, \frac{x}{x-1}\right],
\end{equation}
and noting that the clumping parameter $\nu$ in Kantowski \& Thomas
\shortcite{ka&th00} corresponds to $(\beta -1)/2$. The case of the
cosmological constant is also studied in \cite{kan98,ka&al00,de&al00}.

\section{The luminosity distance in the homogeneous Universe}

\begin{figure}
\epsfxsize=8cm
\centerline{\epsffile{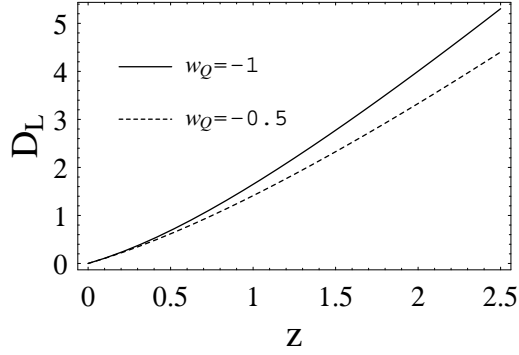}}
\caption{The luminosity distance in a flat and smooth Universe
with $\Omega_M =0.3$. The distance is in units of $c/H_0$.}
\label{dist}
\end{figure}

For the FLRW case ($\alpha_M=1$), Eq.~(\ref{dr4}) is solved by
\begin{equation}
\label{hom1}
D_L(z)=\frac{c}{H_0}(1+z)\int_0 ^z
\frac{1}{\sqrt{\Omega_M(1+z^{'})^3+(1-\Omega_M)
(1+z^{'})^{3(w_Q+1)}}}dz^{'}.
\end{equation}
This expression is equivalent to Eq.~(\ref{dr10}) when $\beta=1/4$,
\begin{equation}
\label{hom2}
D_L(z)=\frac{c}{H_0}\frac{2(1+z)}{\sqrt{\Omega_M}}
\left\{ \ _2F_1\left[-\frac{1}{6 w_Q},
\frac{1}{2}, 1-\frac{1}{6 w_Q},\frac{\Omega_M-1}{\Omega_M} \right] \right.
\end{equation}
\[ \left.
- \frac{1}{\sqrt{1+z}}\ _2F_1\left[-\frac{1}{6 w_Q},
\frac{1}{2}, 1-\frac{1}{6 w_Q},\frac{\Omega_M-1}{\Omega_M}(1+z)^{3w_Q} \right]\right\}.
\]
Our Eq.~(\ref{hom2}) is equivalent to the expression found by
Bloomfield Torres \& Waga
\shortcite{bl&wa96}(see also Giovi \& Amendola
\shortcite{gi&al00}). In the Einstein-de Sitter case ($\Omega_M =1$ or $w_Q
\rightarrow 0$), Eq.~(\ref{hom2}) reduces to
\begin{equation}
\label{hom3}
D_L(z)=\frac{c}{H_0}2(1+z)
\left( 1 - \frac{1}{\sqrt{1+z}} \right),
\end{equation}
as can also be seen directly by solving the integral in
Eq.~(\ref{hom1}). Figure~(\ref{dist}) plots the luminosity distance
for different equations of state: the distance increases for
decreasing $w_Q$.

\section{The magnification probability distribution function}

The amplification of a source at a given redshift has a statistical
nature. For narrow light-beams, the effect of gravitational lensing
results in the appearance of shear and convergence in images of
distant sources according to the different amount and distribution of
matter along different lines of sight. So, gravitational lensing
increases the level of errors in the Hubble diagram
\cite{kan+al95,kan98,frie97,wan+al97,holz98,ho+wa98,metc99,barb00,po+ma00}.
In the framework of the on average FLRW Universe, we can account for
this effect by considering a direction dependent smoothness parameter
$\alpha_M$. Now, $\alpha_M$ represents the effective fraction of
matter density in the beam connecting the observer and the source and
depends on the distribution of matter in the beam \cite{wan99}; values
of $\alpha_M$ greater than $1$ account for amplification effects.

There is an unique mapping between the magnification $\mu$ of a
standard candle at redshift $z$ and the direction-dependent smoothness
parameter at $z$ \cite{wan99}. According to Eq. (\ref{eq:1}), the
magnification $\mu$ of the source with respect to the maximum
empty-beam case ($\alpha_M =0$) is
\begin{equation}
\label{pdf1}
\mu = \left[
\frac{D_L (\alpha_M =0)}{D_L(\alpha_M)}\right]^2.
\end{equation}
Once derived the magnification (that is, once found the distance by
integrating the null-geodesic equation or using ray-tracing techniques
along a line of sight) of a source at epoch $z$, the corresponding
smoothness parameter is determined in comparison with the DR distance:
the solution of the DR equation for that constant value of $\alpha_M$
matches, at redshift $z$, that given value of the distance
\cite{tomi98,wan99}.

The shape of the magnification pdf depends on the redshift of the
source, on the cosmological parameters and on the nature of the DM.
The dark matter can be classified according to its clustering
properties \cite{me+si99,se+ho99,mor+al01}: microscopic DM consists of
weakly interacting massive particles (WIMPs), such as neutralinos
\cite{gu+zy95,gur+al97} and clumps on galaxy halo-scales; macroscopic
DM consists of compact objects, such as massive compact halo objects
(MACHOs) or primordial black holes.

According to N-body simulations of large scale structures in cold dark
matter models, galactic halos are expected to contain a large number
of small substructures besides their overall profile. However, this
type of small-scale structure does not act as a compact object and
only clumps of galaxy-size contribute appreciably to the lensing
\cite{mor+al01}.

In the framework of the on average FLRW models, the $\mu$-pdf is
characterized by some general features with no regard to the nature of
the DM. Under the assumption that the area of a sphere at redshift $z$
centred on the observer is not affected by the mass distribution, the
photon number conservation implies that the mean apparent magnitude of
a source at $z$ is identical to the FLRW value \cite{wein76,sef},
\begin{equation}
\label{pdf2}
\langle \mu \rangle=\mu_{FL} \equiv \mu (\alpha_M =1) >1.
\end{equation}
Since the matter is clumped, most of the narrow light-beams from
distant sources do not intersect any matter along the line of sight
resulting in a dimming of the image with respect to the filled-beam
case: the mode of the pdf, $\mu_{peak}$, is biased towards the empty
beam value,
\begin{equation}
\label{pdf3}
\mu_{peak} < \langle \mu \rangle .
\end{equation}
The third feature is a tail towards large amplifications which
preserves the mean. In terms of the magnification relative to the
mean,
\begin{equation}
\label{pdf4}
\delta \mu \equiv \left[ \frac{D_L (\alpha_M =0)}{D_L(\alpha_M )}\right]^2 - \left[ \frac{D_L (\alpha_M =0)}{D_L(\alpha_M =1)}\right]^2,
\end{equation}
the $\delta \mu$-pdf has the mean at $\delta \mu =0$, the peak value
at $\delta
\mu_{peak} <0$ and a long tail for $\delta \mu >0$, with no regard to the source
redshift and to the cosmological parameters. It follows from these
very general considerations that a simple way to characterize the pdf
is to consider the parameter $\Delta \mu$, defined as the difference
in amplification between the mean FLRW value and the magnification in
the empty beam case ($\alpha_M =0$), $\Delta \mu \equiv
-\delta \mu (\alpha_M
=0)$. When $\Delta \mu$ increases, the mode value moves towards greater
de-magnification: to preserve the total probability and the mean
value, the pdf must both reduce its maximum and enlarge its high
amplification tail. From the properties of the angular diameter
distance in a clumpy Universe \cite{ser+al01}, it follows that $\Delta
\mu$ increases with the redshift of the source and with dark energy
with large negative pressure. So, the dispersion in the $\mu$-pdf due
to gravitational lensing increases with $z$ and it is maximum for the
case of the cosmological constant (see also Bergstr{\"{o}}m et al.
\shortcite{bor+al00}): quintessence with $w_Q>-1$ reduces the bias
towards large de-amplifications of the peak value of the pdf,
partially attenuating the effect of the clumpiness.

\subsection{Lensing by microscopic dark matter}

The gravitational lensing effect by large-scale structure on the
apparent luminosity of distant sources in the Universe has been
studied either with N-body simulations
\cite{wan+al97,tomi98,barb00,bar+al00,jai+al00} or with the
integration of the geodesic deviation equation \cite{ho+wa98,bor+al00}
in a Universe filled with either isothermal spheres or
Navarro-Frenk-White profiles \cite{nav+al95,nav+al96}.

\begin{figure}
\epsfxsize=8cm
\centerline{\epsffile{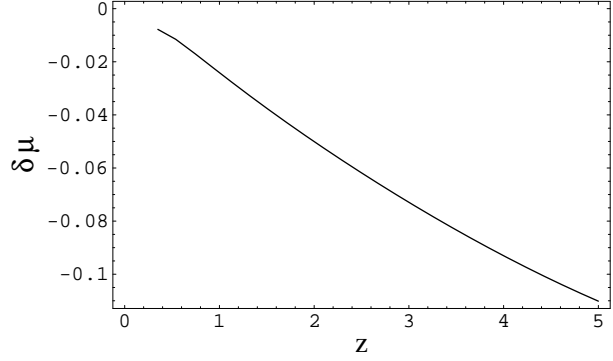}}
\caption{The magnification relative to the mean calculated for the peak
value of the $\alpha_M$-pdf as found in Wang (1999). It is $
\Omega_M=0.4$, $\Omega_Q = 0.6$ and $w_Q = -1$.}
\label{delta_m_smooth}
\end{figure}

The $\mu$-pdf for the smoothly distributed DM is characterized by two
main trends with increasing redshift: an increase in the dispersion
and an increasing gaussianity. As we look back to earlier times, the
Universe becomes smoother on average and lines of sight become more
filled in with matter: light bundles intersect more independent
regions along their paths and the resulting $\mu$-pdf approaches a
gaussian by the central limit theorem \cite{se+ho99,wan99}. The
corresponding $\alpha_M$-pdf also becomes symmetric but it reduces its
dispersion and its mode goes to the filled-beam value
\cite{tomi98,wan99,bar+al00}. The trends in dispersions in the
$\mu$-pdf and $\alpha_M$-pdf are opposite since, with increasing
redshift, a large variation in the distance corresponds to a small
variation in the smoothness parameter \cite{ser+al01}. Wambsganss et
al. \shortcite{wan+al97} used the ray-tracing method for large-scale
simulations in a cold dark matter Universe, normalized to the first
year COBE data with $\Omega_M
=0.4$, $\Omega_Q=0.6$, $w_Q=-1$, with a spatial resolution on small scales of
the order of the size of a halo, to derive the $\mu$-pdf at different
redshifts. Wang
\shortcite{wan99} was able to find empirical formulae for the fitting of
the $\mu$-pdf and of the corresponding $\alpha_M$-pdf,
\begin{equation}
\label{pdf5}
p_{\mu}(\mu, z)=p_{\alpha_M}(\alpha_M,z)\left| \frac{\partial
\alpha_M}{\partial
\mu}\right|=p_{\alpha_M}(\alpha_M,z)\frac{D_A(\alpha_M =0)}{2\mu^{3/2}}\left| \frac{\partial
D_A}{\partial \alpha_M}\right|^{-1}.
\end{equation}
As noted by Tomita \shortcite{tomi98}, the angular diameter distance
depends on $\alpha_M$ linearly for $0 \leq z \stackrel{<}{\sim} 5$
and, with high precision, we can approximate
\begin{equation}
\label{pdf6}
\frac{\partial D_A}{\partial \alpha_M}\simeq D_A(\alpha_M =1)-D_A(\alpha_M =0).
\end{equation}
In Fig.~(\ref{delta_m_smooth}), we plot the $\delta \mu$ corresponding
to the mode of the $\alpha_M$-pdf (as plotted in figure (2b) in Wang
\shortcite{wan99}) as a function of the redshift: while the mode value of
the $\alpha_M$-pdf goes to the filled-beam value for increasing
redshift, the variation in magnification with respect to the FLRW mean
increases; that is, the bias increases with $z$.

\begin{figure}
\epsfxsize=8cm
\centerline{\epsffile{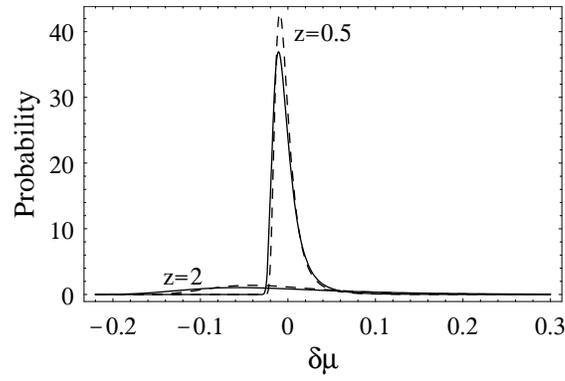}}
\caption{The  amplification pdf as a function of $\delta \mu$, the magnification relative to the
mean, for microscopic DM. The sharply peaked line are for $z=0.5$, the
smoother ones for $z=2$. Solid and dashed lines correspond,
respectively, to $w_Q=-1$ and $-1/2$. It is $\Omega_M=0.4$ and
$\Omega_Q=0.6$. Solid and dashed lines have the same matter
distribution but different cosmological backgrounds.}
\label{mu_pdf_smooth}
\end{figure}

To study the role of the quintessence in the magnification dispersion
of standard candles, we consider the same matter content, that is the
same $\alpha_M$-pdf \cite{wan99}, for different equations of state.
Models with different cosmological parameters produce, in general,
different $\alpha_M$-pdf, predictable by numerical simulations; but,
to consider the influence of the dark energy on the $\mu$-pdf, it
suffices to use the same matter distribution in Eq.~(\ref{pdf5}). This
is equivalent to assume that the dependence on quintessence enters
Eq.~(\ref{pdf5}) through the angular diameter distances and that the
effect on $p_{\alpha_M}$ is of the second order. So, for analytical
convenience, we can use the same $p_{\alpha_M}$ derived in Wang
\shortcite{wan99} for several cosmological models with the same $\Omega_M$
but different equations of state. In Fig.~(\ref{mu_pdf_smooth}), the
$\mu$-pdf is plotted for two source redshifts and for two different
equations of state: the $\mu$-pdf becomes more and more symmetric with
$z$ and the dark energy reduces both the dispersion and the bias.

The effect of gravitational lensing by large-scale structure affects
significantly the determination of the cosmological parameters from
observations of standard candles. Observed SNe Ia represent individual
sources at each redshift and do not sample evenly the probability
distribution: at a fixed redshift, we will observe the mode value of
the distribution and not the mean one \cite{wan+al97,barb00}. For
$\Omega_M =0.4$, $w_Q=-1$ and $z=1$, the mode is $\mu_{peak}=1.14$ and
the magnification values above and below which $97.5\%$ of all of the
lines of sight fall are $\mu_{low}=1.11$ and $\mu_{high}=1.28$. This
dispersion induces uncertainties in determining $\Omega_M$ and the
equation of state. Assuming a flat Universe with cosmological
constant, a Universe with $\Omega_M=0.4$ will be interpreted as a
model with $\Omega_M=0.42^{+0.03}_{-0.11}$ only because of the
gravitational lensing noise. Here and in what follows, the error bars
represent 2-$\sigma$ limits. With the constraint of $\Omega_M=0.4$, a
cosmological constant might be interpreted as dark energy with $w_Q
<-0.84$.

For a flat Universe with $\Omega_M=0.4$ and $w_Q=-0.5$, at $z=1$ it is
$\mu_{peak}=1.11$, $\mu_{low}=1.09$ and $\mu_{high}=1.23$. With the
constraint $w_Q=-0.5$, we should estimate
$\Omega_M=0.43^{+0.05}_{-0.18}$; assuming $\Omega_M=0.4$, it is
$w_Q=-0.46^{+0.05}_{-0.24}$.

Although the lensing dispersion is reduced in a quintessence
cosmology, the errors induced on the cosmological parameters increase.
The reason is that in this models the luminosity distance is less
sensitive to the cosmology \cite{ser+al01}

\subsection{Lensing by compact objects}
The effect of gravitational lensing is maximum when the matter in the
Universe consists of point masses \cite{ho+wa98}; as seen above, this
case is not included in the small-scale structures in the microscopic
DM \cite{mor+al01}. The universal fraction of macroscopic DM is still
unknown. Direct searches for MACHOs in the Milky Way have been
performed by the MACHO and EROS collaborations through microlensing
surveys. According to the MACHO group \cite{alc+al00}, the most likely
halo fraction in form of compact objects with a mass in the range
$0.1-1$ M$_\odot$ is of about $20\%$; the EROS collaboration
\cite{las+al00} has set a $95\%$ confidence limit that objects less
than $1$ M$_\odot$ contribute less than $40\%$ of the dark halo.
However, the average cosmological fraction in macroscopic DM could be
significantly different from these local estimates.

The properties of the $\mu$-pdf are essentially independent of both
the mass spectrum of the lenses (this statement is strictly true for
point sources \cite{sc+we88}) and the clustering properties of the
point masses, provided that the clustering is spherically symmetric
\cite{ho+wa98}. The dispersion in luminosity of standard candles is
non-gaussian, sharply peaked at the empty beam value and has a long
tail towards large magnifications falling as $\mu^{-3}$
\cite{pacz86,rauc91,ho+wa98}, caused by small impact parameter lines
of sight near the compact objects; so, its second moment is
logarithmically divergent and the law of large numbers fails: if
strongly lensed events are removed from the data sample, a bias will
be introduced towards smaller apparent luminousities \cite{ho+wa98}.

A comparative analysis of the $\mu$-pdf in the case of either
microscopic DM or compact objects has put in evidence two main
differences: the high magnification tail is larger for macroscopic DM
and the mode of the distribution is nearer the average value in the
case of lensing by large-scale structures \cite{se+ho99,mor+al01}.

The $\mu$-pdf in a Universe filled with a uniform comoving density of
compact objects depends on a single parameter, the mean magnification
$\langle \mu \rangle$ \cite{rauc91,se+ho99}. Based on Monte-Carlo
simulations, Rauch \shortcite{rauc91} gives the fitting formula
\begin{equation}
\label{pdf7}
p(\mu) \propto \left[ \frac{1-e^{b(\mu -1)}}{\mu^2-1}\right]^{3/2},
\end{equation}
where the parameter $b$ is related to the mean magnification by
$b=247{\rm exp}\left[
-22.3(1-\langle \mu \rangle^{-1/2})\right]$. The approximation holds for $\langle \mu
\rangle^{-1/2} \stackrel{>}{\sim}0.8$, a condition verified up to $z \sim 2$ in a
Universe with low matter density, with no regard to the equation of
state $w_Q$.

\begin{figure}
\epsfxsize=8cm
\centerline{\epsffile{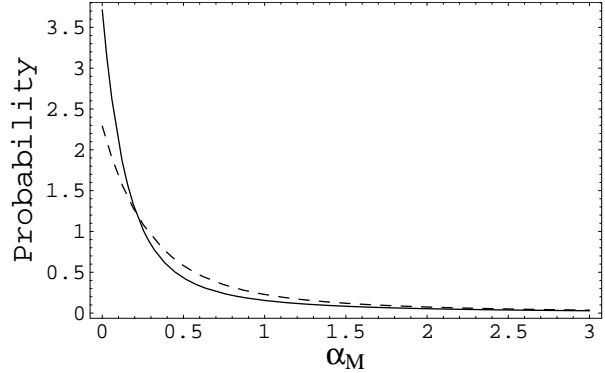}}
\caption{The $\alpha_M$-pdf for macroscopic dark matter. Solid and dashed lines
correspond respectively to $z=0.8$ and $1.5$. It is $\Omega_M=0.3$,
$\Omega_Q=0.7$ and $w_Q=-1$.}
\label{al_pdf_compact}
\end{figure}

The $\alpha_M$-pdf corresponding to the distribution in
Eq.~(\ref{pdf7}) is highly non-gaussian, see
Fig.~(\ref{al_pdf_compact}). The pdf decreases monotonically from the
empty beam value to high values of the smoothness parameter. With
increasing redshift, the $\alpha_M$-pdf tends to flatten and the
probability for the filled-beam case and for high values of $\alpha_M$
grows.

\begin{figure}
\epsfxsize=14cm
\centerline{\epsffile{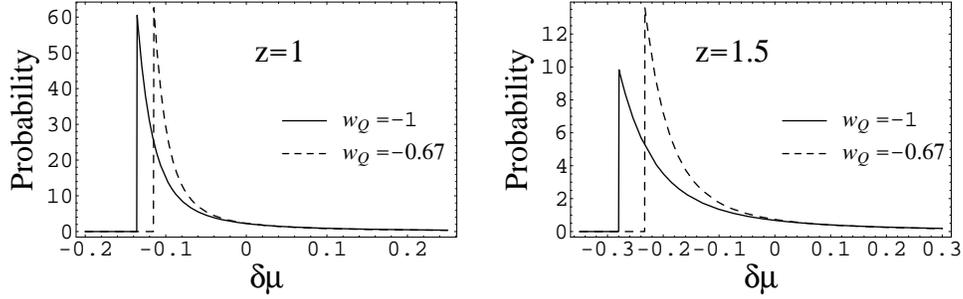}}
\caption{The magnification pdf for macroscopic dark matter as a function
of $\delta \mu$, the magnification relative to the mean. Solid and
dashed lines correspond, respectively, to $w_Q=-1$ and $-2/3$. It is
$\Omega_M=0.3$, $\Omega_Q=0.7$. Left panel: the source redshift is
$z=1$; right panel: it is $z=1.5$}
\label{mu_pdf_compact}
\end{figure}

In Fig.~(\ref{mu_pdf_compact}), we show the $\delta \mu$-pdf for two
source redshifts and for two values of $w_Q$: quintessence with
$w_Q>-1$ reduces the effect of clumpiness. For $z=0.5$, the variation
in the distance modulus from the empty-beam case to the filled-beam
one is $0.033(0.039)$ mag for $w_Q=-1/2(-1)$; for $z=1$, it is
$0.108(0.138)$ mag for $w_Q=-1/2(-1)$; for $z=1.5$, it is
$0.205(0.268)$ mag for $w_Q=-1/2(-1)$. For $z \stackrel{>}{\sim}1$,
the bias towards the empty-beam value can be compared with the
dispersion of $0.17$ mag in the peak magnitudes of SNe Ia after the
application of methods as the ``multi-colour light curve" method
\cite{ri&al96}. The effect of gravitational lensing is of the same
order of magnitude as the other systematic uncertainties that limit
the conclusions on the cosmological parameters based on SNe Ia Hubble
diagram \cite{fi&ri99}. The correlation between host galaxy type and
both luminosity and light-curve shape of the source; interstellar
extinction occurring in the host galaxy and the Milky Way; selection
effects in the comparison of nearby and distant SNe; sample
contamination by SNe that are not SNe Ia can produce changes as large
as $0.1$ mag in the measured luminosities of SNe Ia.

The effect on the estimate of the cosmological parameters of
gravitational lensing by a totally clumped model with only macroscopic
DM is quite dramatic. For a source redshift of $z=1$, a Universe with
$\Omega_M=0.3$ and a cosmological constant can be interpreted as a
model with $\Omega_M=0.42$ and $w_Q=-1$ or as one with $\Omega_M=0.3$
and $w_Q=-0.71$. These systematic errors increase in a quintessence
cosmology with $w_Q
>-1$. For $z=1$, a Universe with $\Omega_M=0.3$ and $w_Q =-2/3$ will be interpreted as
a model with $\Omega_M=0.45$ and $w_Q =-2/3$ or one with
$\Omega_M=0.3$ and $w_Q
=-0.46$.

\begin{figure}
\epsfxsize=8cm
\centerline{\epsffile{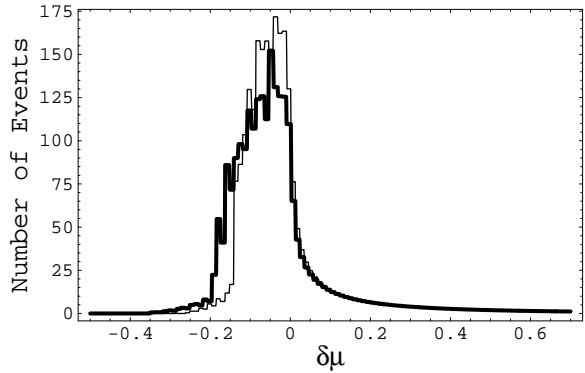}}
\caption{Amplification dispersion relative to the mean due to gravitational lensing
by macroscopic DM for the projected 1-year SNAP sample. Thick and thin
lines correspond, respectively, to $w_Q=-1$ and $-1/2$. It is
$\Omega_M=0.3$, $\Omega_Q=0.7$. Intrinsic dispersion of SN
luminosities is not considered.}
\label{mu_pdf_compact_snap}
\end{figure}

\section{Conclusions}

Observations of SNe Ia are strongly affected by inhomogeneities in the
Universe. For redshifts $z \stackrel{>}{\sim} 1$, the variation in the
distance modulus from a standard flat FLRW model to a clumpy Universe
with the same content of pressureless matter can be considerably
greater than other systematic effects. The effect of amplification
dispersion by gravitational lensing must be accurately considered. The
prospects of future space-born missions, like the SuperNova
Acceleration Probe (SNAP - Http://snap.lbl.gov) and the Next
Generation Space Telescope, of determining properties of the dark
energy have been discussed \cite{gol+al01,we+al01a,we+al01b,ge+ef02}.
According to these studies, SNAP data should only distinguish between
a cosmological constant and quintessence with $w_Q$ relatively far
from $-1$. When SNe observations are combined with an independent
estimate of $\Omega_M$, for example from galaxy clustering
\cite{ver+al01}, the degeneracies among the quintessence models can be
significantly reduced and some constraints on the time evolution of
the equation of state can be put \cite{we+al01b,ge+ef02}. However,
these studies only consider measurement errors and intrinsic
dispersion of the sources, neglecting the systematic and redshift
dependent error induced by gravitational lensing. We have shown how,
also assuming an exact knowledge of $\Omega_M$, in the redshift range
covered by future missions a cosmological constant can be interpreted
as dark energy with $w_Q>-1$. For $\Omega_M=0.4$ and $z=1$, a
$\Lambda$ constant may be interpreted as quintessence with $w_Q
<-0.84$, only due to the lensing by large-scale structure. A fraction
of DM in form of compact objects will make the situation even more
dramatic. So, also with a prior knowledge of the remaining
cosmological parameters, gravitational lensing can make the statements
on the properties of dark energy based on SNe data significantly less
certain.

The effect of inhomogeneities dominates at high redshifts and should
be one of the main systematics in attempting to build the Hubble
diagram with GRBs \cite{nor+al00,rei01a,rei01b,sch+al01}. The physical
origin of GRBs is still uncertain, but recent observations suggest
that they are related to the violent death of massive stars. Under the
hypothesis that GRBs trace the global star formation history of the
Universe, their assumed rate is strongly dependent on the expected
evolution of the star formation rate with the redshift \cite{po+ma01}.
While some scenarios prefer a redshift distribution of the GRB
comoving rate peaked between $z=1$ and $2$, according to other ones
the comoving rate remains roughly constant at $z\stackrel{>}{\sim} 2$
and out to very high redshift \cite{po+ma01}. Furthermore, the lack of
strong lensing events in the fourth BATSE GRBs catalog \cite{hol+al99}
suggests that, at the $95\%$ confidence level, the upper limit to the
average redshift of GRBs is $\stackrel{<}{\sim} 3$ in a flat,
low-matter density Universe with cosmological constant. According to
these considerations, the effect of gravitational lensing would be
really dominant in the Hubble diagram built with GRBs.

As an example, we consider the GRB redshift distribution derived from
a combined analysis of two independent luminosity indicators
\cite{sch+al01}. Examining a sample of 112 GRBs from the BATSE
catalog, Schaefer et al. \shortcite{sch+al01} found redshifts varying
between $0.25$ and $5.9$ with a median of $1.5$. At $z=1.5$,
gravitational lensing by large-scale structures, in a model with
$\Omega_M =0.4$ and $w_Q=-2/3$, induces a magnification distribution
with $\mu_{peak}=1.25$, $\mu_{low}=1.20$ and $\mu_{high}=1.46$.
Assuming $w_Q=-2/3$, we will estimate $\Omega_M=0.43^{+0.05}_{-0.16}$;
assuming $\Omega_M =0.4$, we will estimate $w_Q<-0.51$.

Although the lensing dispersion on the luminosities of standard
candles represents a noise in the determination of the cosmological
parameters, it can also be considered as a probe of the clustering
properties of the DM. Lensing dispersion has been investigated to
search for the presence of compact objects in the Universe
\cite{lin+al88,rauc91,me+si99,se+ho99}. The possibility of determining
the fraction of macroscopic DM using future samples of SNe Ia has also
been explored \cite{mor+al01}. SNAP should intensively observe SNe up
to $z \sim 1.7$. In one year of study, this space-born mission should
be able to discover $\sim 2350$ SNe, most of which in the region $0.5
\stackrel{<}{\sim} z \stackrel{<}{\sim} 1.2$. The discrimination of
models of Universe with different fractions of compact objects is
mainly based on the shift in the peak of the lensing dispersion
\cite{se+ho99,mor+al01}: a shift of $\sim 0.01$ mag in the peak of the
lensing dispersion in the projected SNAP sample towards lower
amplifications corresponds to a growth of $20\%$ in the fraction of
macroscopic DM in a flat Universe with $\Omega_M = 0.3$ and a
cosmological constant (see figure (4) in M\"{o}rtsell et al.
\shortcite{mor+al01}). In Fig.~(\ref{mu_pdf_compact_snap}), we plot the dispersion in
amplification, for the projected redshift distribution of SNe
according to the SNAP proposal, in a Universe with $\Omega_M=0.3$
filled in with macroscopic DM . High de-amplification are preferred in
the case of a cosmological constant, when the maximum of the
distribution is depleted and the mode is shifted away from the mean
with respect to dark energy with $w_Q>-1$. Changing from $w_Q =-1$ to
$w_Q=-1/2$, the peak of the distribution moves for $\sim 0.015$ mag
towards higher amplifications. So, a significant reduction in the
fraction of compact object can be mimed by quintessence with $w_Q
>-1$. Since quintessence reduces the dispersion of gravitational
lensing, it also reduces the ability to distinguish between
microscopic and macroscopic DM from the shape of the amplification
dispersion. Both quintessence and microscopic DM reduce the bias
towards the empty beam value and the high magnification tail and their
effect is of the same order. A Universe with an high fraction of
macroscopic objects can be misleadingly interpreted as one with dark
energy with large negative pressure.

\section*{Acknowledgments}
Authors are indebted to M. Capaccioli, A.A. Marino, C. Rubano and P.
Scudellaro for helpful discussions. They also thank an anonymous
referee for the stimulating reports. MVS also would like to
acknowledge the hospitality of the Universit\`{a} degli Studi di Napoli
``Federico II" and the Osservatorio Astronomico di Capodimonte during
his visit in Napoli. This paper was partially carried out with the
support of ``Cosmion" center, the Russian Fund for Basic Research
(grant N 00-02-16350).

\end{document}